# Design of Ultra-Low Noise Amplifier for Quantum Applications (QLNA)

Ahmad Salmanogli[1] and Vahid Sharif Sirat[2]

[1]Ankara Yildirim Beyazit University, Engineering Faculty, Electrical and Electronic Department, Ankara, Turkey,

salmanogli@aybu.edu.tr

[2]Iran University of Science and Technology, Tehran, Iran

**Abstract:** The present article primarily focuses on the design of an ultra-low-noise amplifier specifically tailored for quantum applications. The circuit design places a significant emphasis on improving the noise figure, as quantum-associated applications require the circuit's noise temperature to be around 0.4 K. This requirement aims to achieve performance comparable to the Josephson Junction amplifier. Although this task presents considerable challenges, the work concentrates on engineering the circuit to minimize mismatch and reflection coefficients, while simultaneously enhancing circuit transconductance. These efforts aim to improve the noise figure as efficiently as possible. The results of this study indicate the possibility of achieving a noise figure of approximately 0.009 dB for a unique circuit design operating at 10 K. In a departure from traditional approaches, this study employs quantum mechanical theory to analyze the circuit comprehensively. By employing quantum theory, the researchers derive relationships that highlight the crucial quantities upon which the circuit design should focus to optimize the noise figure. For example, the circuit's gain power, which depends on the circuit's photonic modes, is theoretically derived and found to affect the noise figure directly. Ultimately, by merging quantum theory with engineering approaches, this study successfully designs a highly efficient circuit that significantly minimizes the noise figure in a quantum application setting.

**Keywords:** ultra-low noise amplifier, noise figure, quantum theory, quantum application

## Introduction:

In recent years, there has been a scientific focus on optimizing classical devices by leveraging quantum phenomena, such as quantum superposition and entanglement [1-5]. This optimization has found widespread applications on large scales, including quantum computers that fundamentally differ from classical ones [4-5], as well as quantum sensors and quantum radars [6-9]. In most of these fields, the Josephson Junction (JJ) plays an indispensable role, thanks to its unique properties, such as superconductivity. For example, the JJ has been used in quantum radar systems to generate entangled microwave photons [9]. Moreover, the JJ can be employed for signal amplification, such as in the Josephson Junction Amplifier (JPA) [9-10], the traveling-wave parametric amplifier (TWPA) [11-12], and the Josephson traveling-wave parametric amplifier (JTWPA) [13]. By utilizing a series combination of JJ elements, such as 1000 JJ or ~2000 JJ, it is possible to amplify very low-level quantum signals. The amplification gain can reach up to 12 dB in the case of JPA [10], and when TWPA or JTWPA is applied with phase matching optimization, it can reach 21 dB [11-13]. One crucial aspect of these amplifiers, making them suitable for various quantum applications, is that they maintain a low noise temperature. The noise temperature of JPA remains around 0.4 K [10], while for TWPA and JTWPA, it is approximately 0.602 K [11-13]. This factor is of particular importance because noise in the system can easily affect entanglement, which is a crucial element in quantum applications. Since amplifying quantum signals is necessary for specific applications, it would be valuable to find elements comparable to JPA or its modified versions.

Recent literature has focused on low-noise amplifiers (LNAs) due to their specific applications, such as quantum computing and astronomy [14-19]. These applications (quantum applications) in contrast with the usual applications require LNA with high sensitivity, very low noise, and high linearity. Moreover, quantum computing and astronomy systems must operate at very low ambient temperatures, necessitating the use of cryogenic LNAs [15-19]. In cryogenic LNAs, the performance of the transistor at extremely low temperatures is a crucial consideration [15-16]. Cryogenic LNAs are essential and often indispensable for branches of science like astronomy [21]. Typically, two types of transistors are utilized in cryogenic LNAs: High Electron Mobility Transistor (HEMT) [19-21] and Silicon-Germanium Heterojunction Bipolar Transistor [20-21]. Notably, HEMT technology remains unaffected by freeze-out at cryogenic temperatures [16-19]. Different types of HEMT, such as TRW InP, Chalmers InP, and Mitsubishi, have been employed in various applications [19]. Consequently, the LNA designed for quantum applications is distinguished by its ability to operate at extremely low temperatures, achieve ultra-low noise figures, consider quantum fluctuation effects, and optimize performance for quantum signals. These features make it a viable alternative or complement to quantum-specific amplifiers in certain quantum computing and sensing applications [15-17].

Knowledge of the mentioned points above has raised the question of whether or how it is possible to amplify the quantum signals with LNA rather than JPA or its modified versions [9-13]. To answer this critical question, some technical comparisons were made between JPA, TWPA, JTWPA [9-13], and LNA (more specifically cryogenic HEMT LNA [14,15]) to clarify the points as follows: The noise temperature is a crucial parameter in quantum applications. Previous research has shown that the noise temperature of a Josephson Junction Amplifier (JPA) operating at 10 mK is around 0.4 K [10], while the noise temperature of a Josephson traveling-wave parametric amplifier (JTWPA) operating at the same temperature is approximately 0.602 K [13]. In comparison, the noise temperature of a cryogenic High Electron Mobility Transistor Low-Noise Amplifier (HEMT LNA) operating at 4.2 K rises to 1.2 K [4]. This comparison reveals that JPA is more efficient than LNA in minimizing noise and its impact on quantum signals. However, the noise temperature of the HEMT LNA remains relatively low and reasonable.

Another critical factor to consider is gain, especially when amplifying very low-amplitude quantum signals. Based on the literature [9-18], it has been found that LNAs can provide significantly higher gain compared to JPAs. Additionally, LNA circuits can be easily engineered to optimize gain, while using 1000 Josephson Junctions in JPA or approximately 2000 unit-cells in JTWPA only result in amplification gains of around 12 dB [10] and 21 dB [13], respectively. The linearity of JPA and LNA can also be compared by defining the linearity of the designed circuit using specific criteria like Input Third-Order Intercept Point (IIP3) or 1 dB Compression Point (P1dB) [24-27]. Linearity is critical because it limits the dynamic range of the circuit, with P1dB being an important parameter. The higher the P1dB, the greater the dynamic range the circuit can achieve. JPA and JTWPA have reported P1dB values around -90 dBm [10] and -98 dBm [13], respectively, while a typical LNA can have a P1dB on the order of -50 dBm. Entering the nonlinearity region can cause power to transfer from the fundamental harmonic to higher-order harmonics. The larger dynamic range of JTWPA leads to increased fabrication complexity [13]. The comparison between JPA and LNA suggests that cryogenic LNAs can be beneficial in quantum computing if two critical factors, namely the noise figure (NF) and power consumption, are optimized. For example, an LNA with a very low NF (e.g., <0.01 dB) and minimal power consumption (e.g., <100 uW) can serve as a pre-amplifier for qubit readout in quantum processors [22]. Based on the aforementioned points, the goal of this work is to design an LNA operating at 10 K, aiming to achieve the smallest possible noise temperature while providing high gain and improved linearity compared to JPAs. With these specifications, the designed LNA can be effectively applied in quantum-associated applications.

## Theory and Backgrounds

### A. *Structure description*

The PCB layout of the initial version of the designed circuit (two-stage LNA) and the related substrate are presented in Fig. 1. The selection of a substrate is a complex decision that involves a trade-off between various electrical, thermal, and mechanical factors. Designers carefully evaluate these parameters to choose a substrate that meets the specific requirements of the circuit and its intended application. The schematic shows only the circuit's passive elements (pink color) containing the Transmission lines (TLs). Achieving the sizes of TLs involves a careful consideration of the electrical properties of the substrate, the characteristic impedance, wavelength, frequency, signal integrity requirements, manufacturing constraints, and design goals. The sizes are often determined through a combination of theoretical analysis, simulation, and practical considerations. Moreover, the active and lumped elements are schematically put on the layout for a better presentation. The lumped components, such as capacitors and resistors, were used to maintain stability and isolate the RF signal from the DC signal. For instance, DC-blocking capacitors are used at the input and output to avoid additional insertion loss. We use the Agilent HEMT ATF-35143 transistor in the design because of appropriate parameters at a cryogenic temperature [17]. The nonlinear model for ATF-35143 with the associated data operating at 10 K is used in the simulation. At cryogenic temperature, the HEMT shows high transconductance and fast transit time resulting in higher gain and higher cut-off frequency. This deviation from the normal mode operations originated from the scattering parameters at a cryogenic temperature [17]. Cooling HEMT produces a substantial improvement in the mobility of electrons and NF.

The designed circuit is biased with $V_{GG}$ = 0.4 V and $V_{DD}$ = 0.47 V, to establish the DC point (graphically illustrated in Appendix B, Fig. $B_1$) by which the drain-source voltage is forced to be as low as possible. The DC point selection is derived from the quantum theory results and it will be discussed in the following. The drain current in the present design is kept at a very low level on the order of 3.2 mA because the high current is a primary source of the shot noise and leads to increased power consumption [17]. The degeneration is used (in the optimized version; Fig. 4a) to obtain better impedance matching to optimize NF. The two-stage LNA circuit is built using Microstrip transmission lines on a 32-mil thickness of RO4003 with $\varepsilon_r$ = 3.55. Conductor Via was also placed in the structure, and a perfect conductor was used in the bottom layer.

### B. *Quantum theory*

In this section, we try to analyze the LNA depicted in Fig. 1 using quantum theory. Using this theory gives some degree of freedom to help any designer to improve the circuit performance. Herein, a theoretical relationship is derived for the power gain of the circuit to engineer NF using quantum theory. In other words, the goal is to find some degree of freedom to manipulate them to improve the circuit's NF. The small signal and simplified model of the LNA illustrated in Fig. 1 are depicted in Fig. $A_2$ (Appendix A). The total Lagrangian of the circuit [30, 31] is given by:

$$L_c = \frac{C_{gs1}}{2}\dot{\varphi}_1^2 - \frac{1}{2L_{g1}}\varphi_1^2 + \frac{C_{gd1}}{2}(\dot{\varphi}_1 - \dot{\varphi}_2)^2 + \frac{C_{in}}{2}(V_{rf} - \dot{\varphi}_1)^2 + \overline{i_n^2}\varphi_1 + \frac{C_{gs2}}{2}\dot{\varphi}_2 - \frac{1}{2L_{d2}}\varphi_2^2 + \frac{C_{gd2}}{2}(\dot{\varphi}_2 - \dot{\varphi}_3)^2 + g_{m1}\dot{\varphi}_1\varphi_2 + \frac{C_{ds3}}{2}\dot{\varphi}_3 - \frac{1}{2L_{d3}}\varphi_3^2 + g_{m2}\dot{\varphi}_2\varphi_3$$

(1)

where $\varphi_1$, $\varphi_2$, and $\varphi_3$ are node fluxes (voltages of the nodes). In Eq. 1, $g_{m1}$ and $g_{m2}$ are the transistor's intrinsic transconductance. $V_{rf}$ and $C_{in}$ are the RF incident wave amplitude and the coupling capacitor between the source and circuit, as shown in Fig. $A_2$ (Appendix A). Additionally, $C_{gs1}$ and $C_{gs2}$ are the capacitance between the gate and source of the transistor, and $C_{gd1}$ and $C_{gd2}$ are the capacitance between the gate and drain of the transistors. Finally, $L_{g1}$, $L_{d2}$, and $L_{d3}$ are the inductance at the gate of the first transistor, the inductance between two transistors, and the inductance of the drain of the second transistor, respectively. Also, $i_n^2$ is the input noise of the system. To simplify the analysis, all noise sources, such as transistor thermal noise ($4KT\gamma g_m B$) and a typical resistor ($R_n$) noise ($4KTR_nB$), are transferred into the input of the circuit and called that $i_n^2$. This is an approximated method that significantly reduces the complexity of the circuit analysis; nonetheless, it keeps the noise effects alive on the circuit [40]. In these equations, B stands for bandwidth. The classical Hamiltonian associated with the circuit is obtained using Legendre transformation $H(\varphi_i,Q_i) = \sum_i (\varphi_i.Q_i) - L_c$, where $Q_i$ is the conjugate variable relating to the coordinate variables $\varphi_i$ satisfying the Poisson bracket $\{ \varphi_i,Q_{i'} \} = \delta_{ii'}$. The conjugate variables are calculated using $Q_k = \partial L_c/\partial(\partial\varphi_k/\partial t)$ [28]. Using the definition listed above, the classical Hamiltonian is derived from Eq. 1 and expressed as:

$$
\begin{aligned}
H_c &= \frac{C_{11}}{2}Q_1^2 + \frac{1}{2L_{g1}}\varphi_1^2 + \frac{C_{22}}{2}Q_2^2 + \frac{1}{2L_{d2}}\varphi_2^2 + \frac{C_{33}}{2}Q_3^2 + \frac{1}{2L_{d3}}\varphi_3^2 \\
&+ \frac{(C_{12}+C_{21})}{2}Q_1Q_2 + \frac{(C_{13}+C_{31})}{2}Q_1Q_3 + \frac{(C_{23}+C_{32})}{2}Q_2Q_3 \\
&+ C_{11}g_{m1}Q_1\varphi_2 + C_{12}g_{m2}Q_1\varphi_3 + C_{21}g_{m1}Q_2\varphi_2 + C_{22}g_{m2}Q_2\varphi_3 + C_{31}g_{m1}Q_2\varphi_2 + C_{32}g_{m2}Q_3\varphi_3 \\
&- C_{in}C_{11}V_{rf}Q_1 - C_{in}C_{21}V_{rf}Q_2 - C_{in}C_{31}V_{rf}Q_3 - \overline{i_n^2}\varphi_1
\end{aligned}
\qquad (2)
$$

where $C_{11}$, $C_{12}$, $C_{13}$, $C_{21}$, $C_{22}$, $C_{23}$, $C_{31}$, $C_{32}$, and $C_{33}$ are constants defined in Appendix A. The first line at the Hamiltonian shows that the circuit fundamentally operates with three simple harmonic oscillators. The second line indicates that the mentioned oscillators are coupled to each other, in which the strength of the coupling relates to the capacitance between the oscillators. The third line in the Hamiltonian reveals that the intrinsic transconductance of the first and second transistors in the circuit changes the Hamiltonian. This is one of the critical points the design focuses on to improve the circuit performance. Finally, the last line shows the effect of the RF incident field on the Hamiltonian and the noise effect.

This work mainly concentrates on minimizing the noise figure of the circuit. Because the noise factor (F) is inversely proportional to the power gain (circuit transconductance) $G_m = I_{out}^2/V_{in}^2$ [40], the study focuses on this quantity and calculates it using the Hamiltonian of the system. In other words, it is necessary to calculate the fluctuation of the circuit transconductance using the Hamiltonian expressed in Eq. 2. Then, by manipulating the critical parameters, one can optimize NF as efficiently as possible. For calculation of the circuit transconductance fluctuation, one needs first to analyze the fluctuation of the output current ($\Delta I_{out}^2 = \langle I_{out}^2\rangle - \langle I_{out}\rangle^2$) and also input node voltage ($\Delta V_{in}^2 = \langle V_{in}^2\rangle - \langle V_{in}\rangle^2$) [28,30] expressed as:

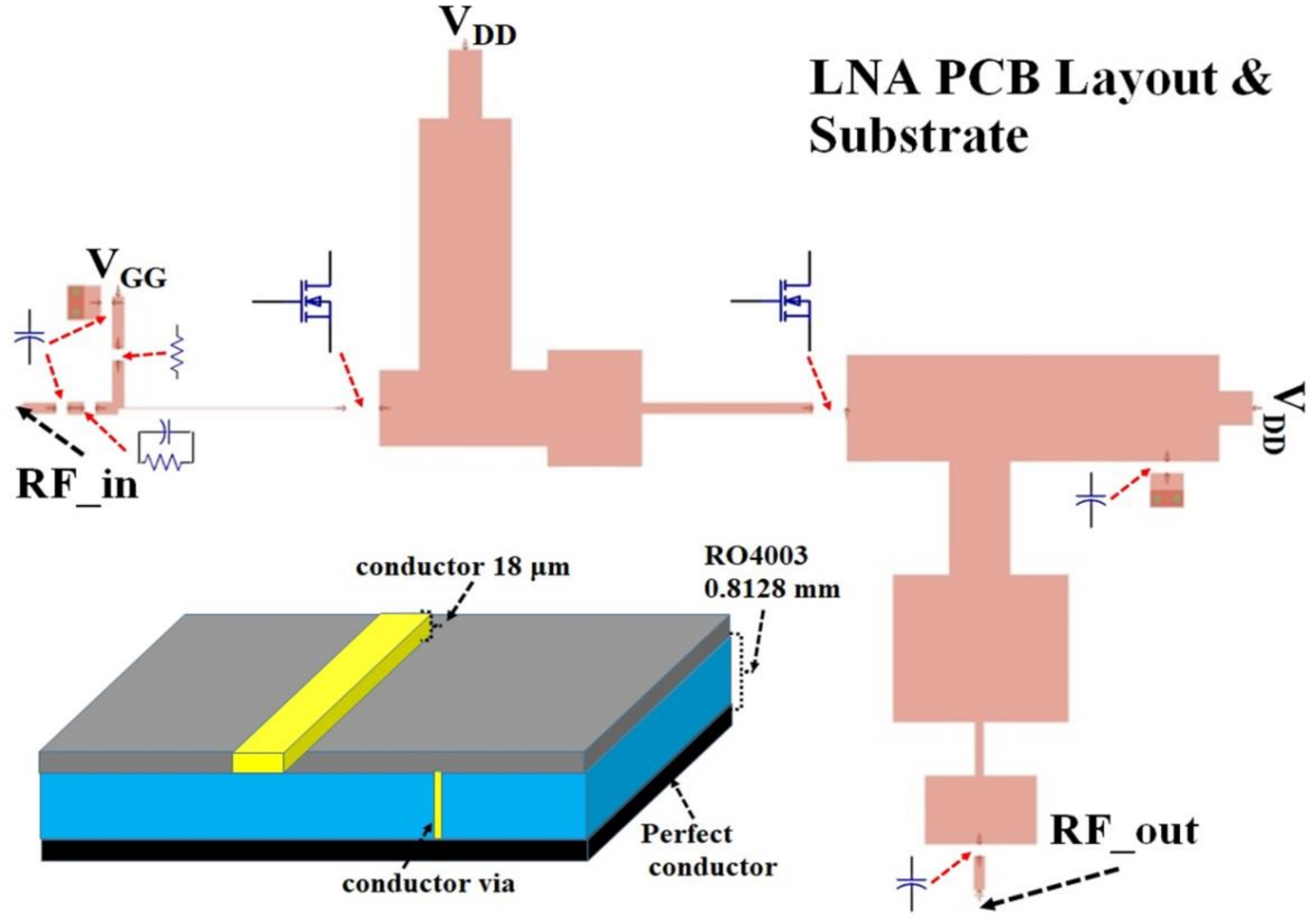

Fig. 1 LNA PCB layout of the circuit and its contributed substrate; the lumped elements used to establish the stability network and also to isolate the RF signals from the DC are depicted in the figures

$$
\begin{aligned}
V_{in} &= \frac{1}{j\hbar}\left[\varphi_1, H_c\right] = C_{11}Q_1 + \frac{(C_{12}+C_{21})}{2}Q_2 + \frac{(C_{13}+C_{31})}{2}Q_3 + C_{11}g_{m1}\varphi_2 + C_{12}g_{m2}\varphi_3 - C_{in}C_{11}V_{rf} \\
I_{out} &= \frac{1}{j\hbar}\left[Q_3, H_c\right] = -C_{12}g_{m2}Q_1 - C_{22}g_{m2}Q_2 - C_{32}g_{m2}Q_3 - \frac{\varphi_3}{L_{d3}}
\end{aligned}
\qquad (3)
$$

Through the calculation of the expectation value of the $\langle Q_i\rangle$, $\langle Q_i^2\rangle$, $\langle \varphi_i\rangle$, and $\langle \varphi_i^2\rangle$ (i = 1,2,3), the circuit transconductance fluctuation is theoretically derived as:

$$\Delta G_m^2 = \frac{\left\{ \frac{(C_{12} g_{m2})^2}{2Z_1}\left(2\overline{n_1}+1\right) + \frac{(C_{22} g_{m2})^2}{2Z_2}\left(2\overline{n_2}+1\right) + \left[ \frac{(C_{32} g_{m2})^2}{2Z_3} + \frac{Z_3}{2L_{d3}^2} \right]\left(2\overline{n_3}+1\right) \right\}}{\left\{ \frac{C_{11}^2}{2Z_1}\left(2\overline{n_1}+1\right) + \left[ \frac{(C_{12}+C_{21})^2}{8Z_2} + \frac{(C_{11} g_{m1})^2 Z_2}{2} \right]\left(2\overline{n_2}+1\right) + \left[ \frac{(C_{13}+C_{31})^2}{8Z_3} + \frac{(C_{12} g_{m2})^2 Z_3}{2} \right]\left(2\overline{n_3}+1\right) \right\}} \quad (4)$$

where $n_1$, $n_2$, and $n_3$ are the expectation value of the oscillators' photon numbers. One can construct the circuit dynamics equation of motions using the Hamiltonian in Eq. 3 to calculate the expectation value of the photon numbers as $n_i = \langle a_i^+ a_i \rangle$, where $a_i^+$ and $a_i$ are the ladder operators [28]. Since the designed circuit interacts with its environment, it can be easily shown that the average photon numbers are strongly affected by the noises generated in the circuit [7, 41]. Eq. 4 clearly shows that the circuit transconductance depends on some parameters of the circuit, more importantly on $g_{m1}$ and $g_{m2}$. This is a key factor that we will use to limit NF. In other words, the transconductance of the stages in the design circuit plays an important role in improving the NF. To show this point, using some assumptions such as $n_1 = 0.1$, $n_2 = 0.56$, $n_3 = 76$, $C_{gs1} = 2.6$ pF, $C_{gs2} = 2.6$ pF, $C_{gd1} = 0.12$ pF, $C_{gd2} = 0.12$ pF, $L_{g1} = 1.1$ nH, $L_{d2} = 2.2$ nH, $L_{d3} = 0.1$ nH, the fluctuation of $G_m$ is modeled and the result is shown in Fig. 2. In Eq. 4, $Z_i$ stands for the oscillator's impedance.

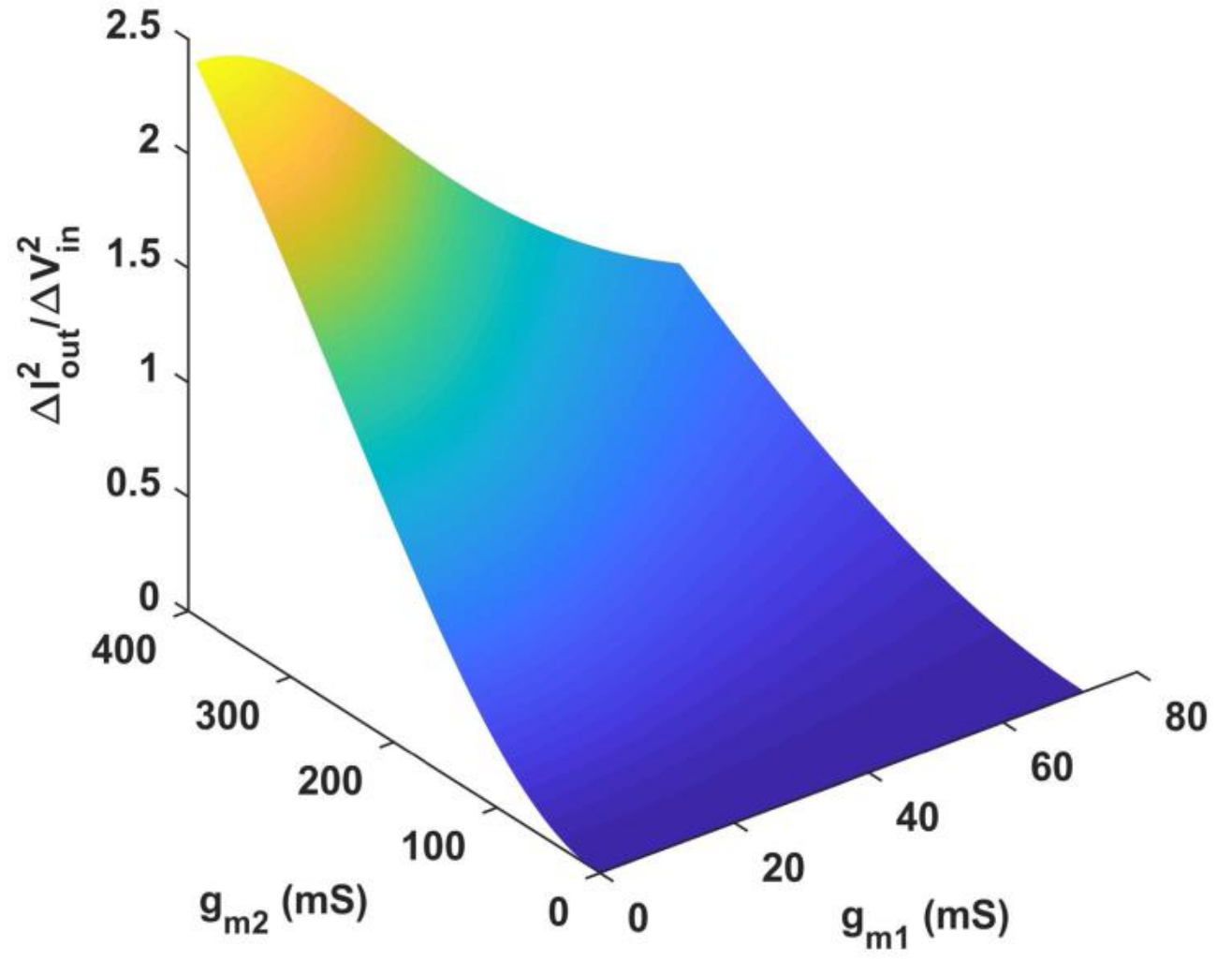

Fig. 2 Circuit transconductance fluctuation vs, $g_{m1}$ (mS) and $g_{m2}$ (mS)

The presented graph shows that the maximum amount of circuit transconductance occurs at the lowest $g_{m1}$ and largest $g_{m2}$. In other words, to minimize NF, which is inversely proportional to the circuit transconductance, it is necessary to keep $g_{m1}$ as small as possible while the simulation shows that there is no limit on $g_{m2}$. The quantum theory simulation, in fact, shows that it is the first stage by which the circuit's NF is restricted. For this reason, the bias current of the first transistor in the circuit should be limited to a low level. Accordingly, the DC bias point of the stages in the amplifier is determined based on the results illustrated in Fig. 2. In addition to the mentioned point, it is shown in the following that the decrease in the bias current strongly improves the dissipation power of the circuit and, in contrast, reduces the speed of the circuit. In other words, this is a trade-off between the critical parameters in the designed LNA, and thus, the problem can be solved by considering the applications for which the LNA is designed.

Matching network designing is one of the important factors by which it will be possible to enhance LNA's power gain, which leads to minimizing NF []. The change in the Microstrip lines' length and width used in LNA led to the difference in the inductance and capacitance in the circuit. Thus, the power gain can be engineered in a way that NF is efficiently minimized. In the classic sense, this is called matching the circuit input with the incident source impedance to reduce the input reflection and circuit output with the load impedance to minimize the output reflection. The standing wave in the circuit is converted to the propagating wave by reducing the reflection coefficients. That is the matter we are looking for it in the design of the LNA. The NF of the circuit is strongly dependent on the input matching network by [17, 30]:

$$F = F_{\min} + \frac{4R_n\,|\Gamma_s - \Gamma_{opt}|^2}{\left(|1-\Gamma_s|^2\right)\left(|1+\Gamma_{opt}|^2\right)} \quad (5)$$

where $F_{min}$, $R_n$, $\Gamma_s$, and $\Gamma_{opt}$ are the minimum noise number depending on the circuit bias point and operating frequency, equivalent noise resistance of the device, source reflection coefficient, and optimum noise source reflection coefficients, respectively. It is clear from Eq. 5 that the selection of $\Gamma_s \sim \Gamma_{opt}$ leads to strongly minimizing the NF in the circuit. Therefore, this is a critical point in which we try to create a partially perfect input and output matching network to strongly minimize NF to reach as close as the quantum limit [21]. In Eq. 5 noise figure is calculated using the classical point of view; however, one can find the other formula in Appendix C, in which the noise factor is theoretically derived using quantum theory; in this formula the expectation value of the oscillator's photon number plays the central role. Thus, using a new point of view gives some degree of freedom by which one can engineer the trade-off in the circuit.

In the next section, the emphasis lies on engineering the input and output matching network used in LNA to optimize NF.

**Results and Discussions**

This section focuses on the analysis and study of specific technical characteristics of the designed Low-Noise Amplifier (LNA). All simulations were conducted using Advanced Design System (ADS) on the PCB layout at an operating temperature of 10 K. The use of electromagnetic (EM) simulation in Fig. 1 and Fig. 4a enhances the accuracy of the analysis. The EM simulation results are presented in Fig. 3 and Fig. 4. Notably, the circuit was designed to be unconditionally stable within the frequency range of 1.0-3.0 GHz (Fig. A3 in Appendix A). Ensuring circuit stability is of utmost importance because increasing gain or minimizing noise figure (NF) at cryogenic temperatures can lead to instability. Traditional techniques such as resistive loading and negative feedback can stabilize the circuit, but negative feedback with capacitance is preferred in this design to mitigate the thermal noise effect of resistors.

Fig. 3a illustrates the NF results within the considered bandwidth. To calculate the NF of the designed circuit, the input and output were terminated with a 50Ω load, and the input and output reflection coefficients were used to determine the noise factor (F) [24-27] (Eq. 5). The design achieves a minimum NF of approximately 0.04 dB within the bandwidth. This value is attained by optimizing the circuit's reflection coefficient. Manipulating the length and width of the circuit's Microstrip lines enables control over power reflections caused by impedance mismatch ($\Gamma_s$ and $\Gamma_{out}$). Minimizing losses induced in the circuit and improving power gain to enhance NF requires reducing impedance mismatch in the design, where Γs should be in the order of $\Gamma_{opt}$. However, the primary focus of the designed circuit is on minimizing NF and power consumption, potentially sacrificing other important features such as bandwidth and gain. This trade-off is necessary for the LNA to be utilized effectively in quantum applications, where a very low NF comparable to JPA's NF is required. Reference literature [10] shows that JPA demonstrates a minimum NF of around 0.007 dB at 10 mK. Comparing the NF of the LNA with JPA reveals the need for further optimization to minimize the circuit's NF. After studying other parameters of the initial design (Fig. 1) and comparing them with JPA's technical characteristics, another LNA was developed to reduce NF and approach the quantum limit significantly. Power dissipation is another crucial feature of the LNA, and the obtained results are presented in Fig. 3b. The maximum power dissipated by the designed circuit within the frequency range is approximately 1.43 mW. This value is reasonable when compared to other LNAs designed for low NF [24-27].

Furthermore, the results of the simulations, particularly power consumption, can be compared with cryogenically operated LNAs [22]. Another critical feature of the LNA is its gain, as illustrated in Fig. 3c, which is approximately 22 dB. The graph demonstrates that the designed LNA operates within the linear region when the RF input power is below -30 dBm. However, when the input power exceeds -25 dBm, the LNA circuit enters the nonlinear region, leading to a significant decrease in gain. This occurs because the input power is distributed among other harmonics, as discussed later. The linearity of the LNA explains the interference of spurious signals with the operating frequencies of the LNA. To delve into this point, the linearity of the LNA is analyzed and depicted in Fig. 3d. The graph compares the fundamental and third-order output power. It is evident from the comparison that below -120 dBm, the power shared with third-order harmonics is negligible, on the order of -250 dB. Furthermore, the third-order intercept point (IIP3) is calculated as a measure of linearity, and the result indicates that the linearity of the designed LNA occurs at approximately -15 dBm. However, for quantum applications such as quantum radars, incident or backscattered photons are typically at a low level, around 0.2-1.2 Photon/Hz. The incident power fed into quantum radar applications is reduced to approximately ~-180 dBm. Therefore, the designed LNA can safely operate within the linear region.

Finally, the 2-tone test (1.6 GHz ± 1 MHz) results of the LNA are shown in Fig. 3e and Fig. 3f. This test aims to determine the extent to which the fundamental power is shared specifically with the third and fifth modes when the device is excited with two separate modes (classically correlated modes). The detuning frequency is selected at around 1 MHz, meaning that the LNA is simultaneously excited with two modes at 1599 MHz and 1601 MHz. Fig. 3e illustrates the intermodulation distortion (IMD) of the third and fifth orders. This criterion quantifies the power difference between the fundamental mode, third-order mode, and fifth-order mode. According to the figure, it is evident that when the input power is below -105 dBm, the corresponding IMD remains below -180 dB, indicating that the IMD is much less than 0.1 Photon/Hz. Therefore, this performance is suitable for quantum radar and other quantum sensors with an average incident power of approximately 1 Photon/Hz (-152 dBm) [9]. Moreover, Fig. 3f displays the spectrum containing the two original modes and other harmonics.

The considerable variation in NF at cryogenic LNAs can be attributed to two main factors: 1. The higher speed electron mobility, which reduces the thermal noise of the channel, and 2. The engineering of the matching network to minimize NF. In this study, the focus is primarily on the latter case, aiming to engineer Γs to control and minimize NF. To achieve this, four LNAs with different techniques were designed and their simulation results are shown in Fig. 4. Fig. 4a illustrates the PCB layout of the cryogenic LNA designed to provide an ultra-low noise level of around 0.01 dB, suitable for quantum applications. In this circuit, negative capacitive feedback and degenerative impedance in the source of the transistors are utilized to optimize impedance matching, in contrast to our initial LNA design depicted in Fig. 1. Fig. 4b presents the EM-simulation result, where the PCB layout has been modified to achieve optimum matching between Γs and Γout.

Four different cryogenic LNA PCB layouts were established, with one schematically shown in Fig. 1 and the optimized version with very low noise depicted in Fig. 4a. For the other two LNAs, we present only the NF and gain results. These four different versions of LNAs were designed to demonstrate the trade-off between gain and NF. The design aimed to minimize NF while maintaining the LNA's stability and partially enhancing the circuit's gain as much as possible. This critical achievement resulted in reducing the LNA's NF to 0.009 dB, as shown in Fig. 4b, while the gain of the circuit was significantly compromised. The inset plot in Fig. 4b demonstrates that the NF fluctuates in the range of 0.009-0.012 dB within the bandwidth of 1.15-3.0 GHz. This indicates that the noise temperature of the designed LNA, schematically depicted in Fig. 4a, varies between 0.6-0.8 K in the considered bandwidth. These results in Fig. 4b can be compared with Fig. 4c, where it is evident that minimizing NF leads to a decrease in the circuit's gain. The presented NF results are comparable to [32]. While the designed LNA with very low NF is suitable for quantum applications, it suffers from a significant reduction in gain. Fig. 4c illustrates the gain of the designed circuits, where S(2,1) scattering parameters determine the forward gain. Consequently, this circuit, as an ultra-low-noise amplifier, can be employed as a pre-amplifier in qubit readout quantum circuits. However, it faces a critical challenge in terms of gain.

It can be argued that the designed LNA, with its excellent linearity, stability, reasonable noise temperature, and modest gain, can partially replace JPA in quantum computing applications, such as qubit readout circuits as a pre-amplifier. This is particularly relevant due to the impedance mismatching that causes additional decoherence in superconducting qubits due to wave backscattering from the first-stage amplifier. Moreover, while the qubits operate at 10 mK, the first-stage amplifier operates at 4.2 K, resulting in a significant thermal mismatch between the two regions. Considering the aforementioned points, if the gain of the LNA designed in Fig. 4a is increased to a comparable level (around 21 dB), it would be possible to replace JPA or JTWPA with the solid-state LNA. Nonetheless, the LNA with low gain but optimized NF of approximately 0.01 dB has the ability to generate nonclassicality between microwave photons [41]. Finally, the results of this study are compared with state-of-the-art LNAs given in Table. I. In addition, the extra technical information about the LNA, such as its Figure of Merit (FOM), is calculated [34-36] and presented in Appendix C (Fig. $C_1$). The calculated results are comparable with [39]. Furthermore, in this appendix, some critical metrics such as gain, nonlinearity, and noise figures of the designed LNA are calculated at different temperatures and compared with each other (Fig. $C_2$). In other words, the effect of the temperature on some essential metrics of LNAs is investigated.

In the following, we tried to apply another precise method to analyze the noise performance of the circuit and calculate the input and output signal-to-noise ratio and then figure out NF. This method is a different approach than the method considered to estimate NF in Fig. 3a and Fig. 4b. The latter approach is based on the impedance matching between input and output; meanwhile, the present method, called the noise analyzing method, applies an input signal, 1-tone or two-tone, to the circuit and calculates the associated signal to noise ratio. This method calculates the fast Fourier transform for each time (0-1000 μsec with a step of 100 nsec). In this approach, 1.6 GHz is typically selected as the center frequency, and the method calculates the detuning frequency about the center frequency and analyzes the parameters. In this method, the input signal with a typical power on the order of 1 Photon/Hz can be applied, meaning that the circuit designed directly senses an input that could be applicable in the quantum realm, following which the signal-to-noise ratio is calculated.

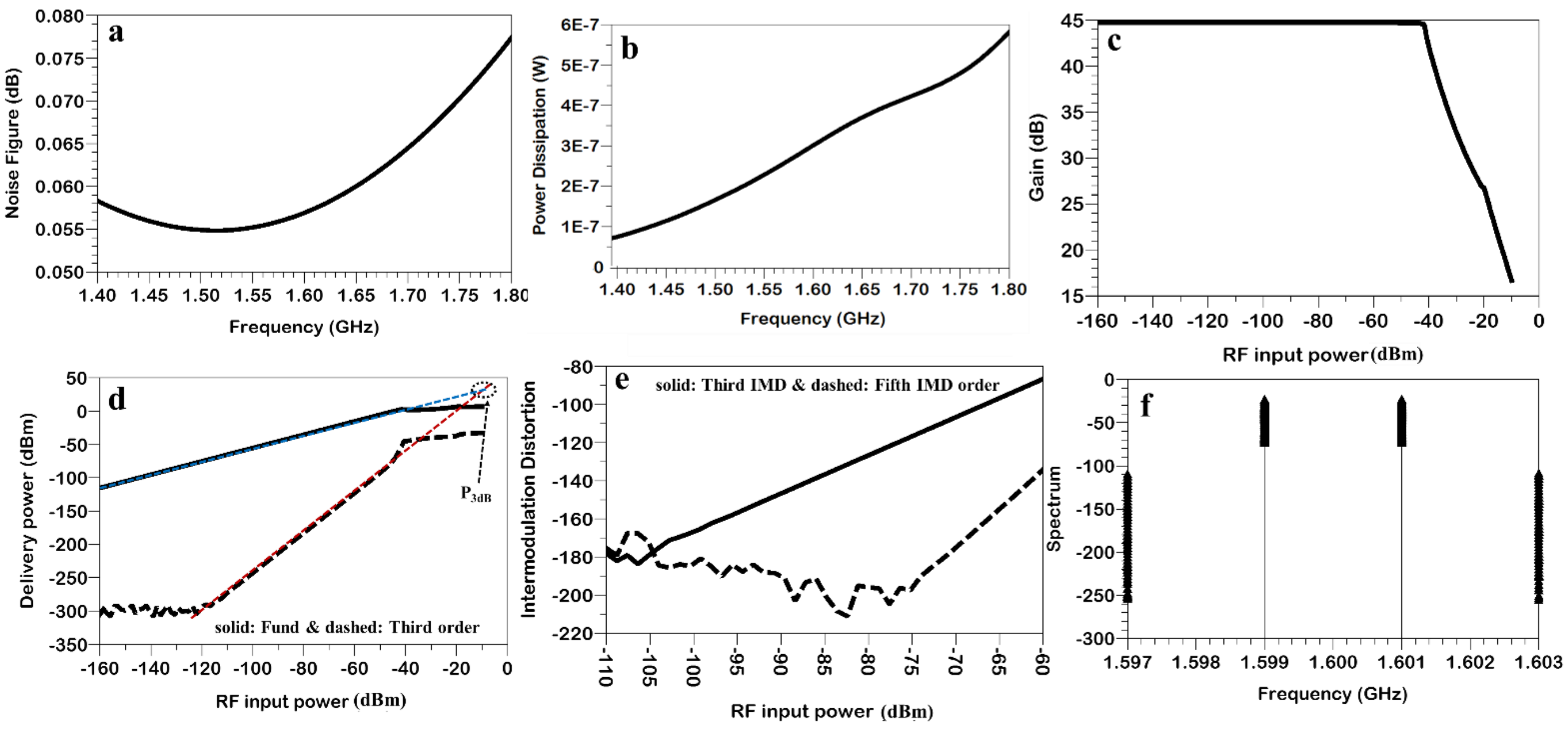

Fig. 3 a) QLNA noise figure (dB) vs. frequency (GHz) b) power dissipation (W) vs. frequency (GHz) for input power -50 dBm, c) gain (dB) vs. RF input power (dBm) for incident frequency 1.6 GHz, d) Fundamental and third-order output power vs. RF input power (dBm) for incident frequency 1.6 GHz, e) Intermodulation distortion vs RF input power (dBm) for incident frequency 1.6 GHz±1 MHz, f) 2-tone test Spectrum for incident frequency 1.6 GHz ±1 MHz.

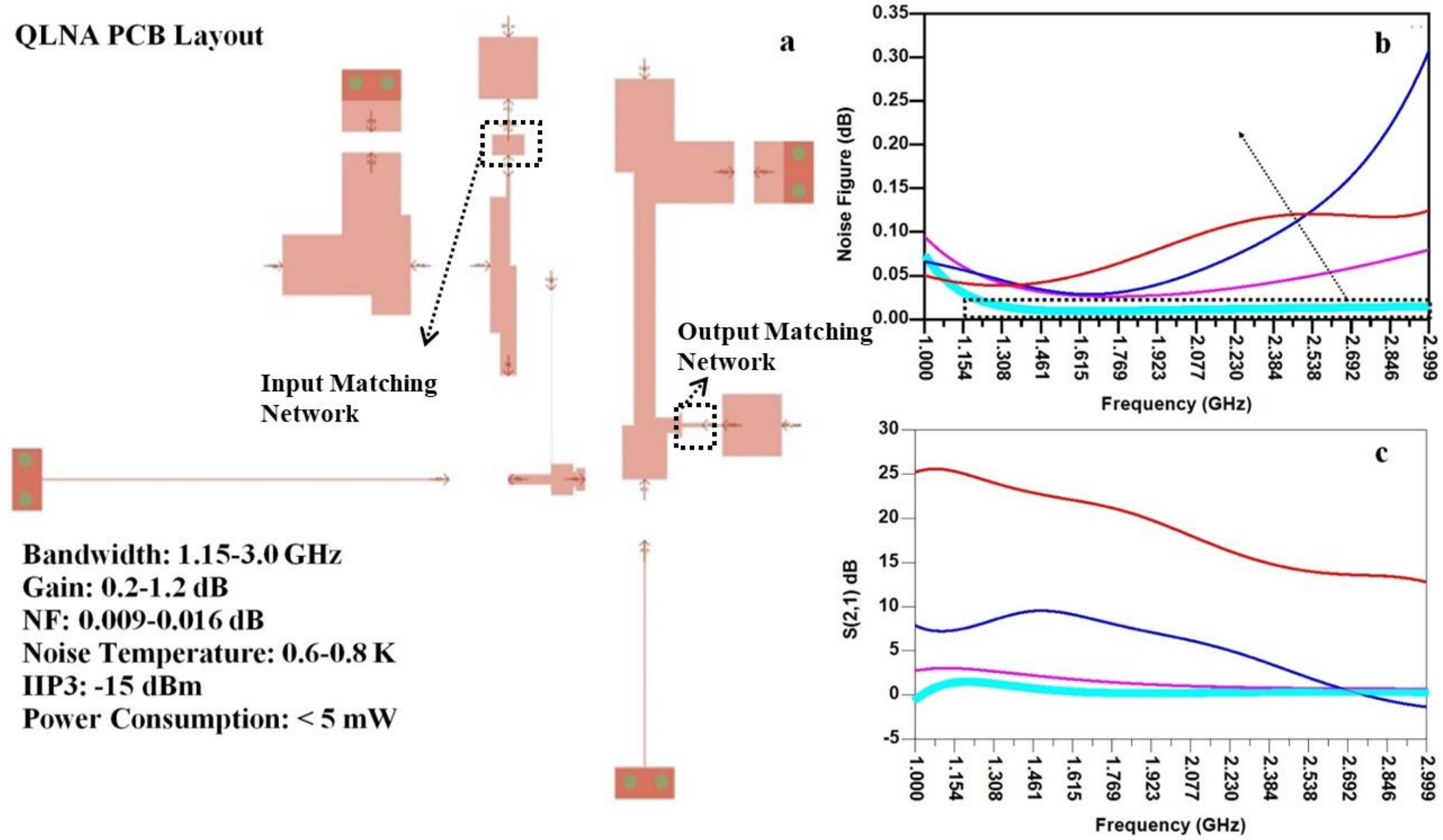

Fig. 4a) Schematic of the PCB layout of the LNA circuit designed to be operated in the quantum applications and its list specifications, b) Noise figure of four different designed LNA, c) the comparison between the gain of the four additional LNA.

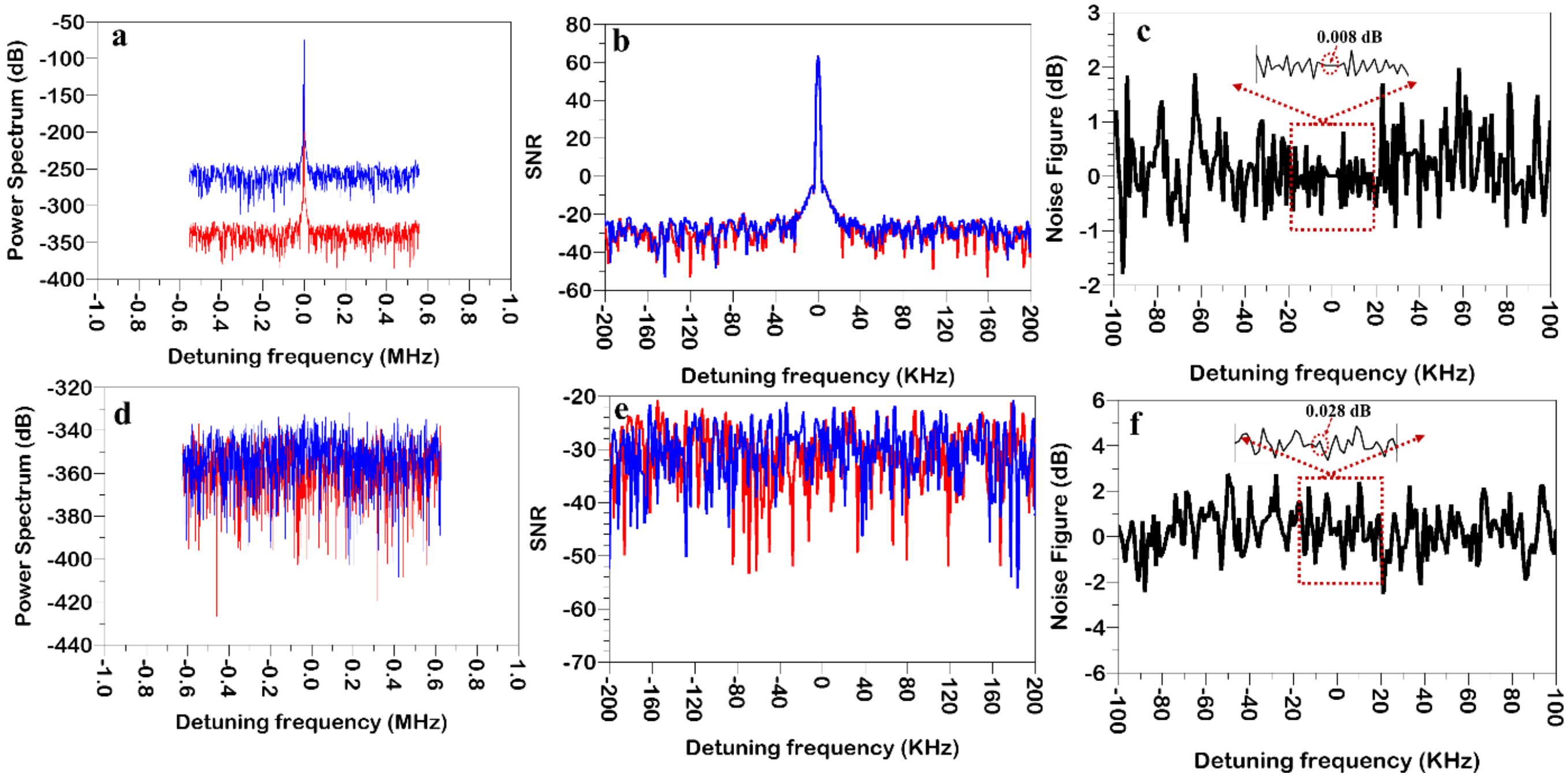

Fig. 5 LNA noise analysis; upper row: 1-tone test, down row: 2-tone test; (a,d) input (red) and output (blue) power spectrum vs. detuning frequency (MHz), (b and e) input (red) and output (blue) SNR vs. detuning frequency (KHz), (c and f) Noise figure (dB) vs. detuning frequency (KHz).

The noise analysis is performed for both 1-tone and 2-tone power input signals. Fig. 5a and Fig. 5d display the input and output power spectra for 1-tone and 2-tone signals, respectively. In the case of 1-tone analysis, the circuit exhibits a positive signal-to-noise ratio (SNR) around the center frequency. Using this graph, the noise factor ($F = SNR_i/SNR_o$), where $SNR_i$ and $SNR_o$ represent the input and output SNR, is calculated [36-37]. Subsequently, NF = 10log(F) is evaluated [36-37], and the NF graph is presented in Fig. 5c. With 1-tone excitation, the noise factor is significantly reduced, reaching approximately 0.008 dB. The corresponding noise temperature associated with this NF is approximately 0.53 K, which is comparable to the noise temperature of JPA [10]. However, it is important to note that generating 1-tone microwave photons is extremely challenging, and in practical applications, such as quantum radars, it is impossible to feed an LNA with a 1-tone frequency signal. In real-world scenarios, thermal noise and other types of noise can impact the signal, leading to a broadened bandwidth of the received signals that the LNA senses. Hence, the designed LNA is excited with 2-tone signals, such as 1.6 GHz ± 2 MHz, and the simulation results are shown in Fig. 5d, Fig. 5e, and Fig. 5f. From Fig. 5e, it is evident that the input and output SNR are significantly affected. Furthermore, Fig. 5e demonstrates that the NF increases to 0.028 dB at the center frequency, corresponding to a noise temperature of approximately 1.88 K. These results indicate that it is possible to design an LNA that can improve the noise temperature to some extent, making it partially comparable to JPA.

Table .1 Comparison with the state-of-the-art LNAs

| | Freq (GHz) | NF (dB) | Gain (dB) | Tem (K) |
|---|---|---|---|---|
| Ref [35] | 2-2.12 | 0.5 | 18 | 77 |
| Ref [22] | 4-8 | 0.05 | 20 | 5 |
| Ref [34] | 0.1-5 | 0.04-0.07 | 28-33 | 15 |
| Ref [33] | 4.6-8.2 | 0.23-0.65 | 39-44 | 4.2 |
| This work | 1.1-3.0 | 0.009-0.01 | 0.1-1.2 | 10 |

**Conclusions:**

This article primarily focuses on the design of an LNA for quantum applications, aiming to achieve performance comparable to that of JPA, which maintains a remarkably low noise temperature of approximately 0.4 K. Four different 2-stage LNAs were designed, with a specific emphasis on minimizing the NF while considering the trade-off between NF, gain, and stability factors. In the initial stage, the designed LNA was analyzed using quantum theory, specifically examining the quantum fluctuation of the circuit transconductance, which is a critical factor affecting NF. The analysis revealed that the first stage (gm1) should be limited to minimize NF in the circuit. Based on this insight, the PCB layout of the circuit was established and subjected to simulation. The results of the electromagnetic (EM) simulation demonstrated that the LNA achieved an impressive gain of approximately 22 dB and exhibited excellent linearity suitable for quantum applications, operating around -20 dB. Additionally, the intermodulation distortion for the LNA, with an RF input power below -150 dBm (suitable for quantum applications), was found to be approximately -180 dB. However, the initial LNA design exhibited a noise figure of approximately 0.04 dB. To optimize the NF in the design, compensative techniques such as negative feedback and degenerative impedance in the transistors' source were employed. By implementing these techniques, the NF of the newly designed LNA was significantly reduced to 0.009 dB, making it suitable for quantum applications and comparable to the noise temperature of JPA. However, the gain of the new design experienced a significant decrease, ranging from 0.21 to 1.2 dB across the bandwidth.

## Appendix

*Appendix A:*

The schematic of the 2-stage LNA is illustrated in Fig. $A_1$. We used this general circuit to generate its PCB layout shown in Fig. 1 and then optimized the circuit for minimum NF. In Fig. $A_2$, the small signal model of the circuit is illustrated. This model is used to quantum mechanically analyze the circuit. To calculate the gain power, this model is used.

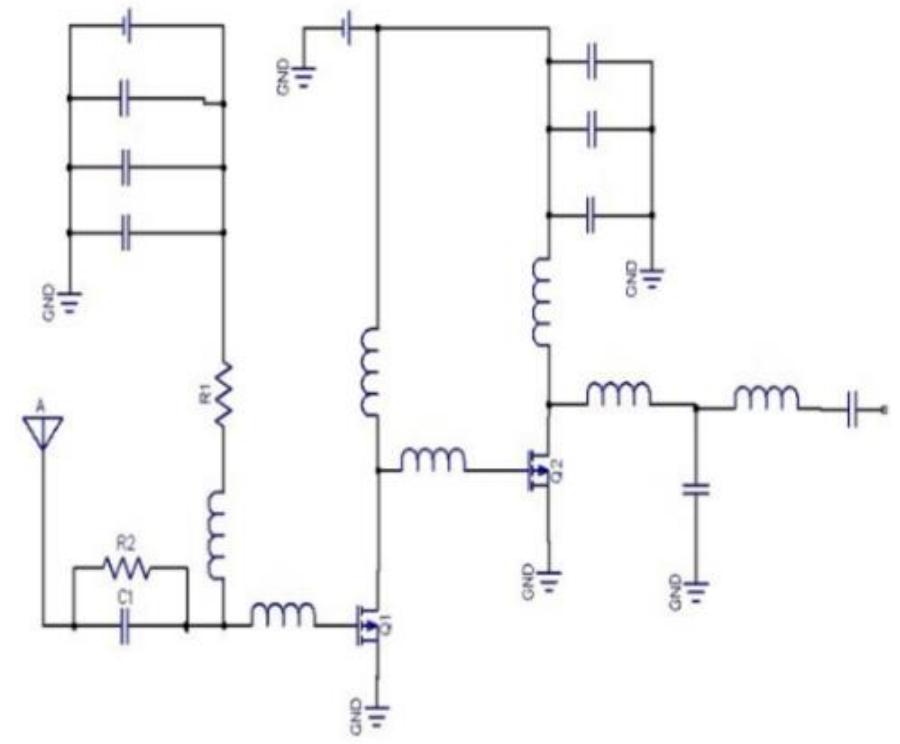

Fig. $A_1$ LNA circuit general schematic

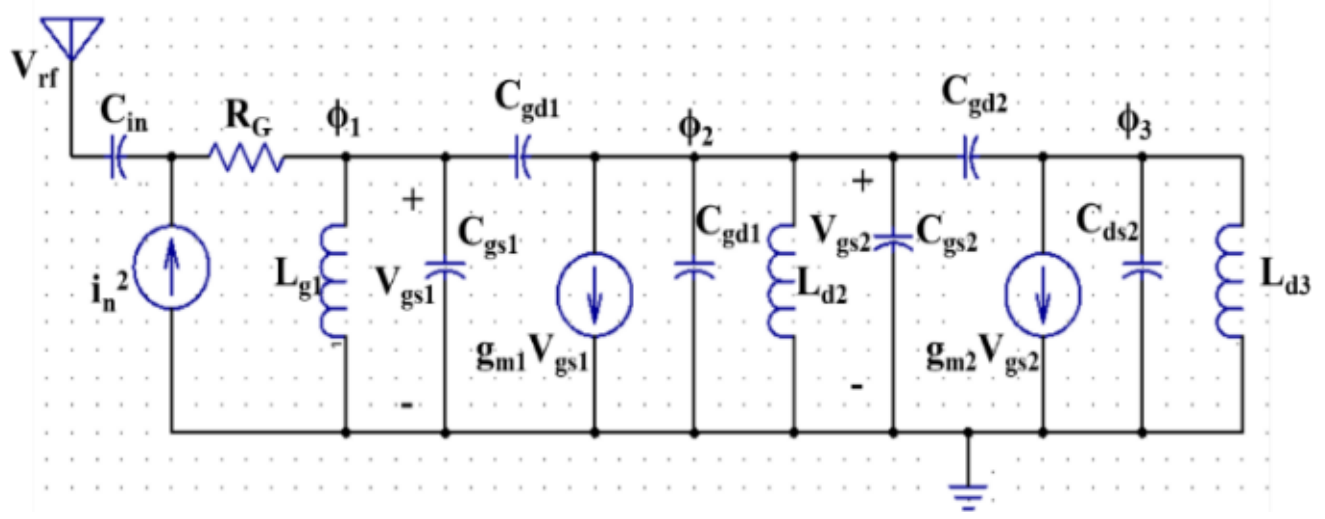

Fig. $A_2$ Small signal model for the designed LNA

The LNA stability factor in the considered bandwidth is illustrated in Fig. $A_3$. It is shown that in the considered bandwidth, the stability factor is greater than 1, meaning that the designed circuit is stable.

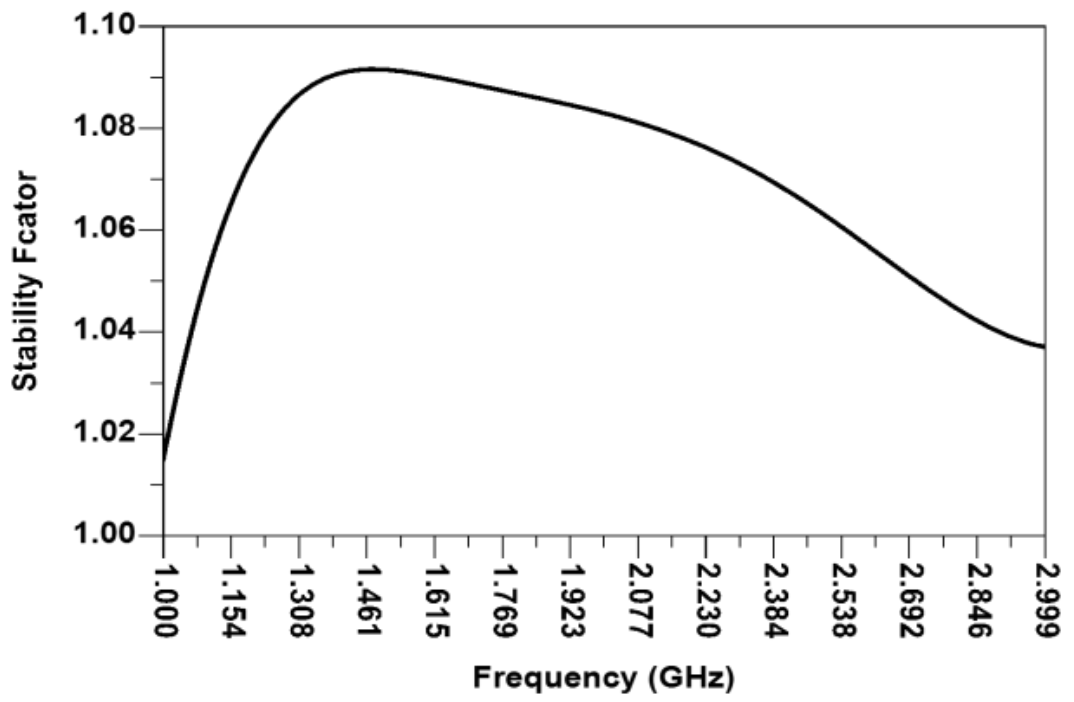

Fig. $A_3$ Stability factor vs. Frequency (GHz)

In the following, we try to clearly express all of the constants used in the equations of the main article. The capacitance matrix for the circuit is given by:

$$\begin{bmatrix} Q_1 \\ Q_2 \\ Q_3 \end{bmatrix} = \begin{bmatrix} C_1 & -C_{gd1} & 0 \\ -C_{gd1} & C_2 & -C_{gd2} \\ & -C_{gd2} & C_3 \end{bmatrix} \begin{bmatrix} \dot{\varphi}_1 \\ \dot{\varphi}_2 \\ \dot{\varphi}_3 \end{bmatrix} + \begin{bmatrix} -C_{in} & 0 & 0 \\ 0 & 0 & 0 \\ 0 & 0 & 0 \end{bmatrix} \begin{bmatrix} V_{rf} \\ 0 \\ 0 \\ 0 \end{bmatrix} + \begin{bmatrix} 0 & g_{m1} & 0 \\ 0 & 0 & g_{m1} \\ & 0 & 0 \end{bmatrix} \begin{bmatrix} \varphi_1 \\ \varphi_2 \\ \varphi_3 \end{bmatrix}, \begin{cases} C_1 = C_{in} + C_{gs1} + C_{gd1} \\ C_2 = C_{gd2} + C_{gs2} + C_{gd1} \\ C_3 = C_{ds3} + C_{gd2} \end{cases} \quad (A_1)$$

Using some algebra, $\partial_t \varphi_i$ can be expressed in terms of $Q_i$ to establish the classic Hamiltonian as:

$$\begin{bmatrix} Q_1 \\ Q_2 \\ Q_3 \end{bmatrix} = \begin{bmatrix} C_{11} & C_{12} & C_{13} \\ C_{21} & C_{22} & C_{23} \\ C_{31} & C_{32} & C_{33} \end{bmatrix} \begin{bmatrix} \dot{\varphi}_1 \\ \dot{\varphi}_2 \\ \dot{\varphi}_3 \end{bmatrix} - \begin{bmatrix} C_{11} & C_{12} & C_{13} \\ C_{21} & C_{22} & C_{23} \\ C_{31} & C_{32} & C_{33} \end{bmatrix} \begin{bmatrix} -C_{in} & 0 & 0 \\ 0 & 0 & 0 \\ 0 & 0 & 0 \end{bmatrix} \begin{bmatrix} V_{rf} \\ 0 \\ 0 \\ 0 \end{bmatrix} - \begin{bmatrix} C_{11} & C_{12} & C_{13} \\ C_{21} & C_{22} & C_{23} \\ C_{31} & C_{32} & C_{33} \end{bmatrix} \begin{bmatrix} 0 & g_{m1} & 0 \\ 0 & 0 & g_{m1} \\ & 0 & 0 \end{bmatrix} \begin{bmatrix} \varphi_1 \\ \varphi_2 \\ \varphi_3 \end{bmatrix} \quad (A_2)$$

$$\begin{bmatrix} C_1 & -C_{gd1} & 0 \\ -C_{gd1} & C_2 & -C_{gd2} \\ & -C_{gd2} & C_3 \end{bmatrix}^{(-1)} \equiv \begin{bmatrix} C_{11} & C_{12} & C_{13} \\ C_{21} & C_{22} & C_{23} \\ C_{31} & C_{32} & C_{33} \end{bmatrix}$$

*Appendix B:*

This appendix shows the circuit's LNA bias characteristics and scattering parameters (small signal analysis) at the selected bias points ($V_{DS}$ = 0.4 V, $V_{GS}$ = 0.4 V, and $I_{DS}$ = 10 mA) in Fig. $B_1$ and Fig. $B_2$, respectively. The LNA bias characteristic is shown in Fig. $B_1$, in which the operating point is indicated with a dashed arrow in the figure. Moreover, the designed LNA small signal parameters are illustrated in Fig. $B_2$. The curve $S_{21}$ shows that the circuit's gain is very low, contributing to the LNA circuit design in which we attempted to minimize the noise figure. In addition, the comparison between NF and $NF_{min}$ is demonstrated in Fig. $B_3$.

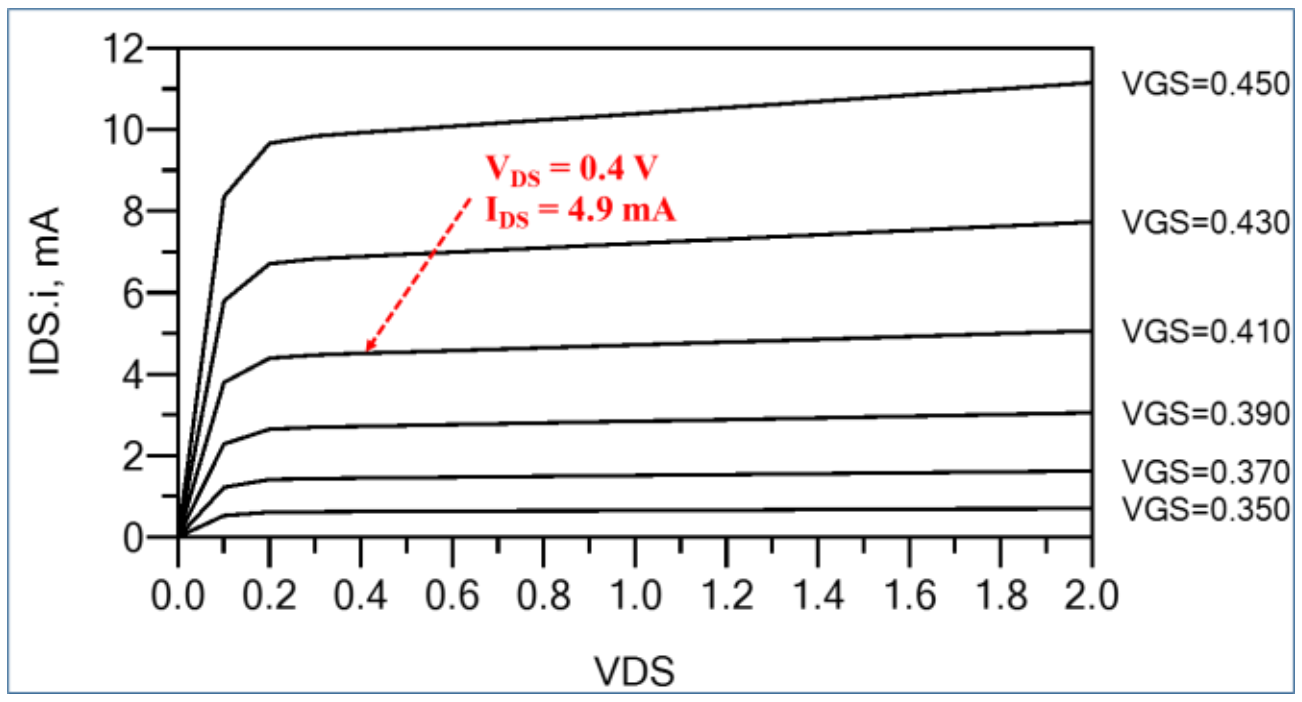

Fig. $B_1$ LNA bias characteristics

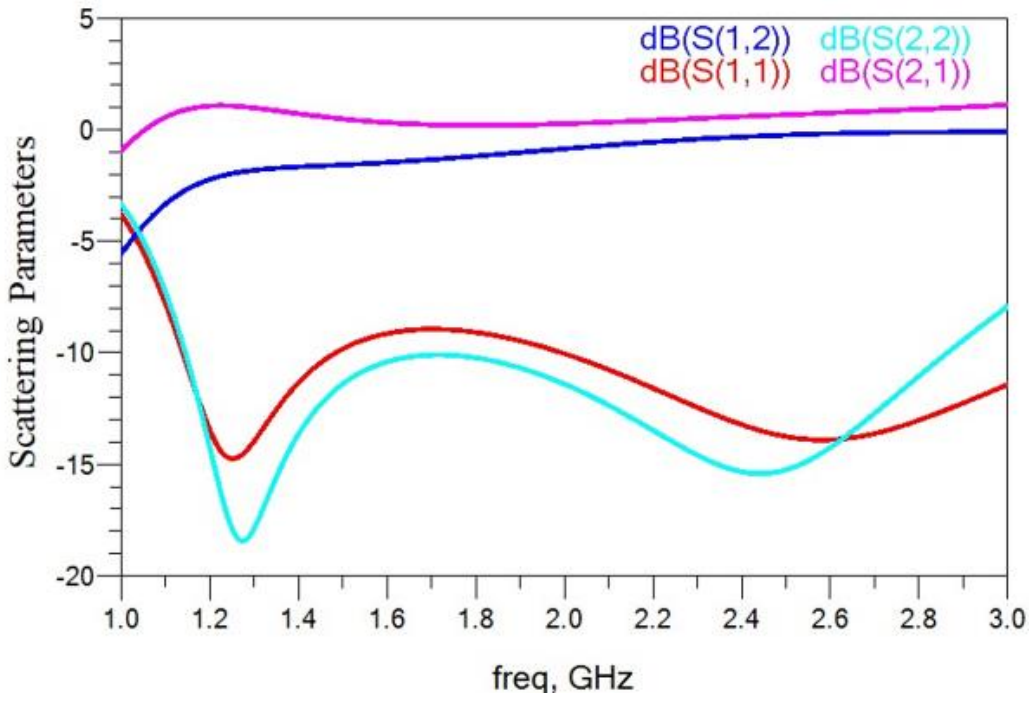

Fig. $B_2$ LNA Scattering parameters

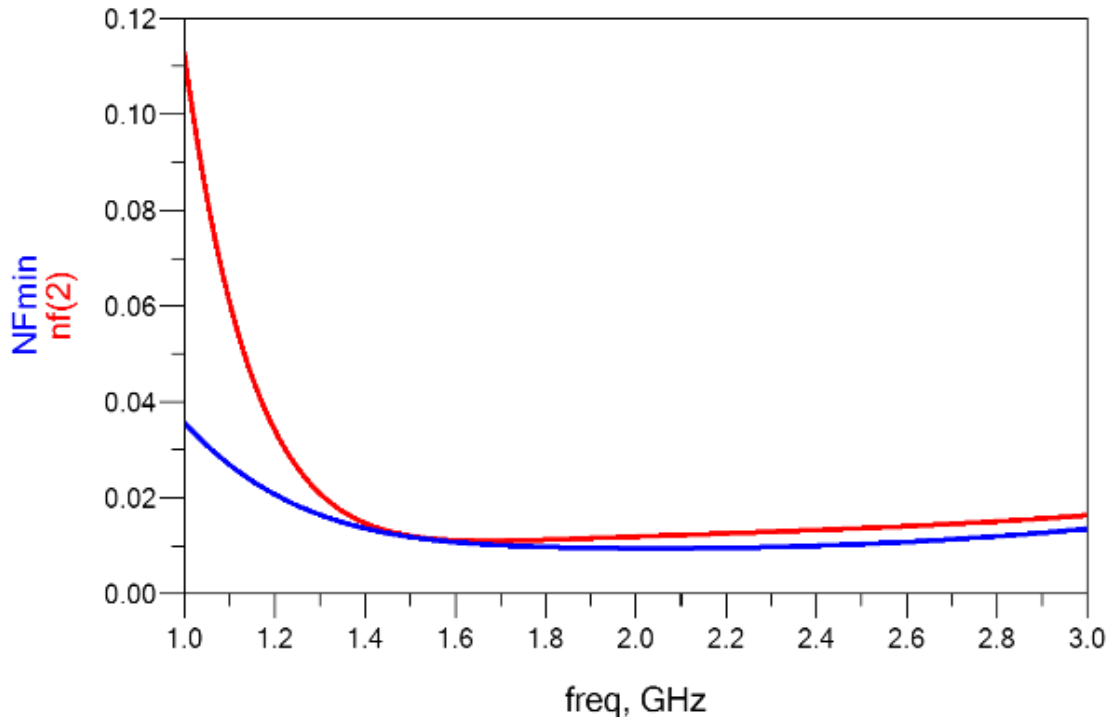

Fig. $B_3$ NF and $NF_{min}$ vs frequency (GHz)

Appendix C:

*Noise number (F):*

According to the Friis equation, the overall noise figure of cascaded stages is dominated by the first amplification stage [36-37]. Generally, the noise performance of a circuit is characterized by the noise factor, F, defined as the ratio of the input signal-to-noise ratio and the output signal-to-noise ratio given by:

$$F = \frac{N_{out}}{\Delta G_m . N_{in}} \tag{C_1}$$

where $\Delta G_m^2$ is the power gain, which is theoretically derived in Eq. 4 of the main paper; in addition, $N_{in}$ is the source noise power, and $N_{out}$ is the output load noise power. $\Delta G_m$ is classically introduced as $|S_{21}|^2/(1-|S_{22}|^2)$, which depends on the gain of the circuit and output reflection coefficients [36-37]. Therefore, using $C_1$ and the result of Eq. 4 of the main paper, the noise number is calculated as:

$$F \; \alpha \; \frac{\left\{ \frac{C_{11}^2}{2Z_1}\left(2\bar{n}_1+1\right)+\left[\frac{\left(C_{12}+C_{21}\right)^2}{8Z_2}+\frac{\left(C_{11}g_{m1}\right)^2 Z_2}{2}\right]\left(2\bar{n}_2+1\right)+\left[\frac{\left(C_{13}+C_{31}\right)^2}{8Z_3}+\frac{\left(C_{12}g_{m2}\right)^2 Z_3}{2}\right]\left(2\bar{n}_3+1\right)\right\}}{\left\{\frac{\left(C_{12}g_{m2}\right)^2}{2Z_1}\left(2\bar{n}_1+1\right)+\frac{\left(C_{22}g_{m2}\right)^2}{2Z_2}\left(2\bar{n}_2+1\right)+\left[\frac{\left(C_{32}g_{m2}\right)^2}{2Z_3}+\frac{Z_3}{2L_{d3}^2}\right]\left(2\bar{n}_3+1\right)\right\}} \tag{C_2}$$

Here, Eq. $C_2$ shows that the noise number is strongly manipulated by the photon numbers and the contributed coefficients related to the circuit elements. Thus, using quantum theory analysis opens a new view of the point by which one can easily manipulate the noise figure just by focusing on the photon numbers of the oscillators of the circuits. It is too different from the classical view that only concentrates on the circuit's elements.

*3rd-order intercept point ($IIP_3$) theory*

Here, in this section, we try to shortly introduce $IIP_3$ as an essential phenomenon [24-25],[36-37] in a nonlinear system. When two unmodulated sinusoidal signals with different frequencies (slightly separated) are applied to the input of a nonlinear system, then some other components with varying frequencies of the input signals appear at the output. For instance, if one supposes input signals as $V_{in} = A_1\cos(\omega_1 t) + A_2\cos(\omega_2 t)$, then the output of the circuit designed containing nonlinearity can be introduced as $V_{out} = C_1V_{in} + C_2V_{in}^2+C_3V_{in}^3$, which is issued due to the nonlinearity of the transistor in the system. Therefore, by substituting $V_{in}$ into $V_{out}$, the output voltage becomes:

$$V_{out} = C_1\left\{A_1\cos\left(\omega_1 t\right)+A_2\cos\left(\omega_2 t\right)\right\}+C_2\left\{A_1\cos\left(\omega_1 t\right)+A_2\cos\left(\omega_2 t\right)\right\}^2+C_3\left\{A_1\cos\left(\omega_1 t\right)+A_2\cos\left(\omega_2 t\right)\right\}^3 \tag{C_3}$$

Using some algebra (breaking the square and cube formula) and finally using trigonometric relationships eventuates the linear and third-order components at the output simplified as:

$$V_{out} \; \alpha \left[C_1A_1\cos\left(\omega_1 t\right), C_1A_2\cos\left(\omega_2 t\right)\right]+\left[\frac{3C_3A_1^2A_2}{4}\cos\left(\left\{2\omega_1-\omega_2\right\}t\right), \frac{3C_3A_2^2A_1}{4}\cos\left(\left\{2\omega_2-\omega_1\right\}t\right)\right] \tag{C_4}$$

$IIP_3$ point (intercept point) is where the two output power curves, linear component and third order component, intercept with each other. Generally, the intercept point is a characteristic of a system's linearity.

*LNA's figure of merit*

The FOM of the designed LNA is calculated [37-39], and the result is illustrated in Fig. $C_1$. It is shown in this figure that FOM becomes maximum around 1616 GHz, at which the noise figure becomes maximum (the result is shown in Fig. 4b of the main article). In addition, the results illustrated in Fig. $C_1$ are comparable with the LNA's related FOM in [39].

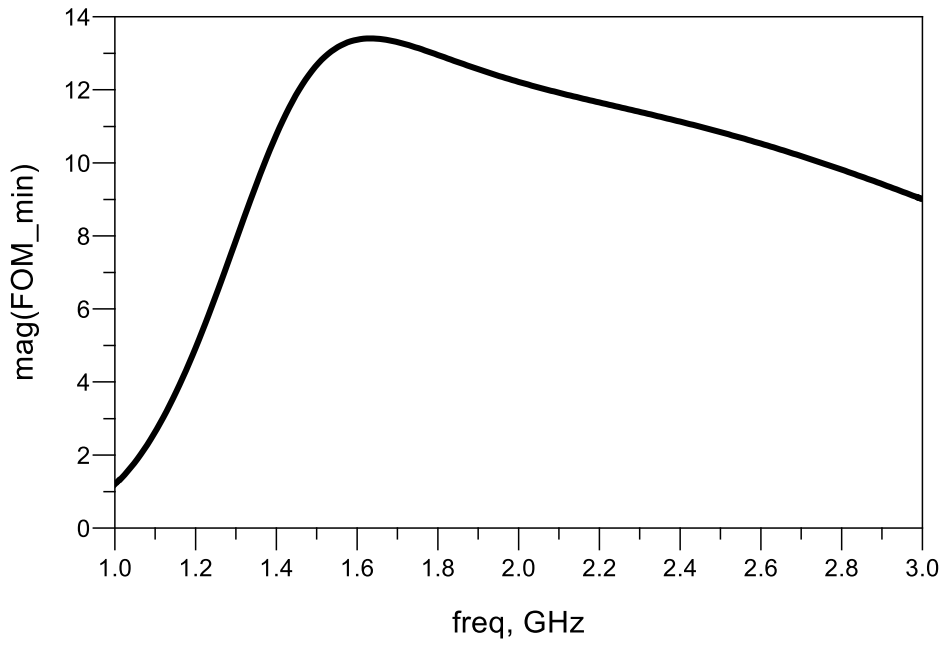

*Fig. $C_1$ Figure of merit (FOM) for the designed LNA*

*LNA's important metrics at different temperatures:*

The LNA's (the optimum design that contains the minimum noise figure and is discussed in Fig. 4 of the main paper) essential metrics such as NF, Gain, and $IIP_3$ are calculated at different temperatures, and the results are illustrated in Fig. $C_2$. The results show that the noise figure is strongly increased by increasing temperature, and the circuit causes nonlinearity behavior to initiate so earlier. Finally, the gain of the circuit is decreased.

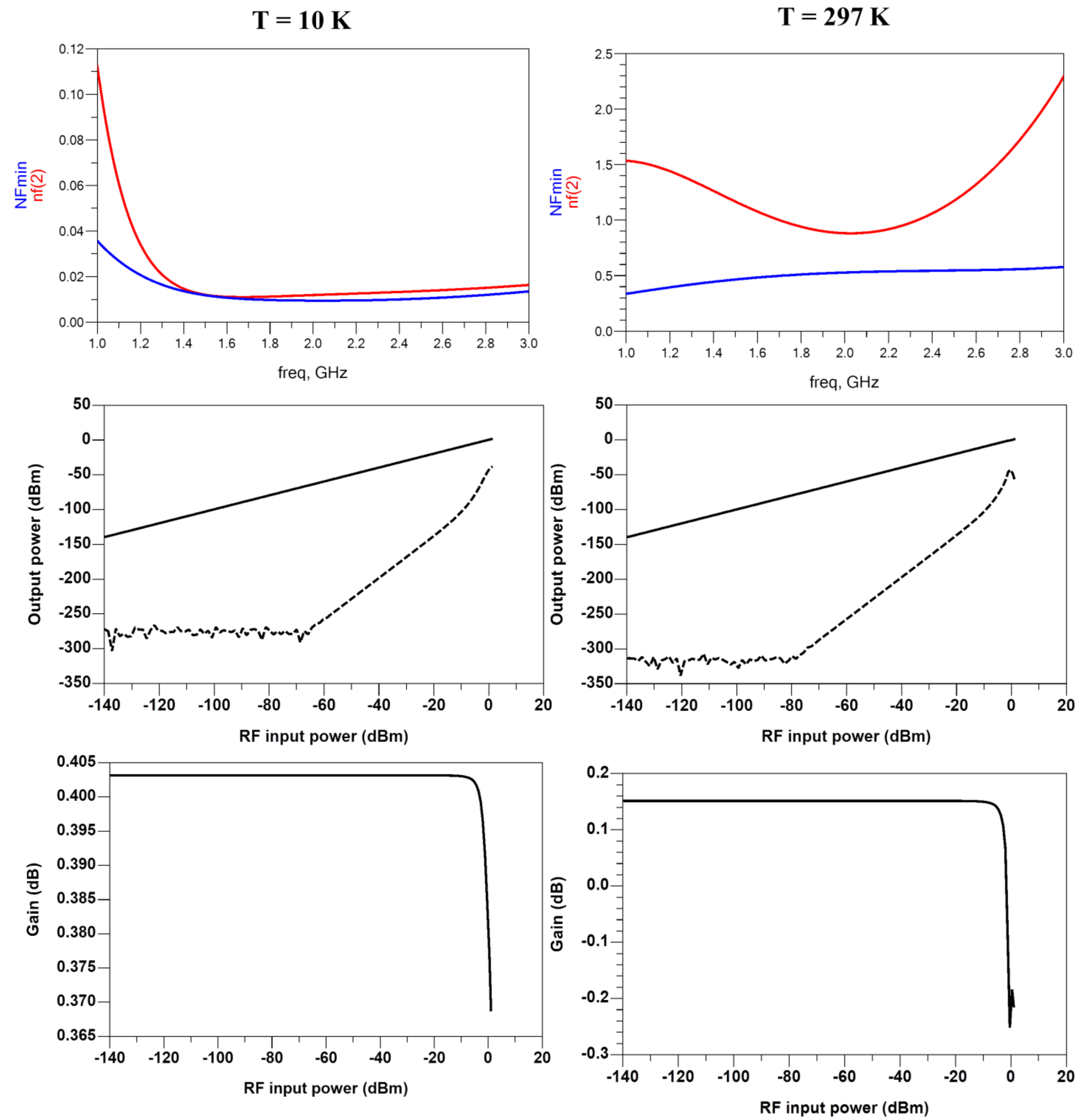

Fig. $C_2$ LNA's essential metrics such as NF (up row), $IIP_3$ (middle row), and Gain (down the row), were calculated at different temperatures at 10 K (left column) and 297 K (right column).

Conflict of interest: There is no conflict of interest for this work.

Ethics approval and consent to participate: I confirm that this work is original and has been neither published nor is currently under consideration for publication elsewhere.

Consent for publication: The author of this study gives the publisher the permission of the author to publish the work.

Availability of data and materials: There are no datasets generated for this work; however, the code generated during the current study is available from the corresponding author upon reasonable request.

Competing interests: There are no competing interests.

Funding: There is no funding for this work.

Authors' contributions: All of the studies have been done by Ahmad Salmanogli.